\documentclass[11pt,a4,twosided]{article}
\usepackage{cqmoriond,graphicx}
\usepackage{natbib}
\usepackage{bibentry}
\usepackage{hyperref}
\usepackage[french,english]{babel}
\usepackage{amsfonts}

\newcommand{\gev}{\hbox{ GeV}}

\newcommand{\mev}{\hbox{ MeV}}

\newcommand{\tev}{\hbox{ TeV}}

\newcommand{\fm}{\hbox{ fm}}

\newcommand{\mb}{\hbox{ mb}}

\newcommand{\cfrac}[2]{\textstyle \frac{#1}{#2}}

\def\ltap{\raisebox{-.4ex}{\rlap{$\sim$}} \raisebox{.4ex}{$<$}}

\def\be{\begin{equation}}
\def\ee{\end{equation}}
\def\bea{\begin{eqnarray}}
\def\eea{\end{eqnarray}}

\begin{document}
\begin{flushright}
\textsf{FERMILAB--CONF--08--430--T}
\end{flushright}
\vspace*{4cm}
\title{THEORETICAL PERSPECTIVES\\ XLIII RENCONTRES DE MORIOND---QCD}

\author{ Chris QUIGG }

\address{Theoretical Physics Department, \\ Fermi National Accelerator Laboratory,\\
P.O. Box 500, Batavia, Illinois 60510 USA \\ and \\ Institut f\"{u}r Theoretische Teilchenphysik \\ 
Universit\"{a}t Karlsruhe\\ D-76128 Karlsruhe, Germany}

\maketitle
\vspace*{-5mm}
\begin{figure}[h]
\centerline{\includegraphics[width=35mm]{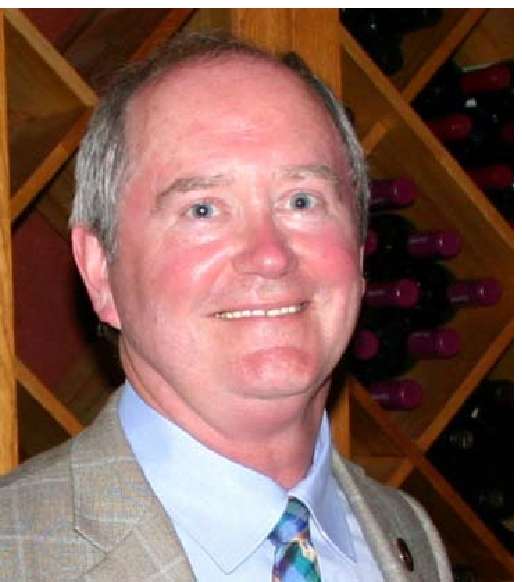}}
\end{figure}

\abstracts{I offer a brief summary, with commentary, of theoretical contributions to Moriond QCD 2008.} 

\section{Asymptotic Freedom and the Strong Coupling Parameter $\alpha_s$}
Quantum chromodynamics has been with us for thirty-five years as the theory of the strong interactions, and the news of this meeting is that the pace of important developments---both in experiment and in theory---is impressively high. The magical property of QCD is that it is an asymptotically free theory, so that the ``strong'' interactions observed at low energy scales yield at higher scales to interaction strengths that can be controlled in perturbation theory. In leading logarithmic approximation, the strong coupling $\alpha_s$ runs according to the simple equation
\begin{equation}
\frac{1}{\alpha_s(Q)} = \frac{1}{\alpha_s(\mu)} + \frac{(33 - 2n_f)}{6\pi}\ln\left(\frac{Q}{\mu}\right)\;,
\end{equation}
where $n_f$ is the number of quark flavors active at the scale $Q$. So long as $n_f \le 16$, the coefficient of the logarithmic term is positive, and the coupling strength diminishes at high energy scales. In the world we have explored, $n_f \le 6$.

\begin{figure}[t!]
\centerline{\includegraphics[width=8.0cm]{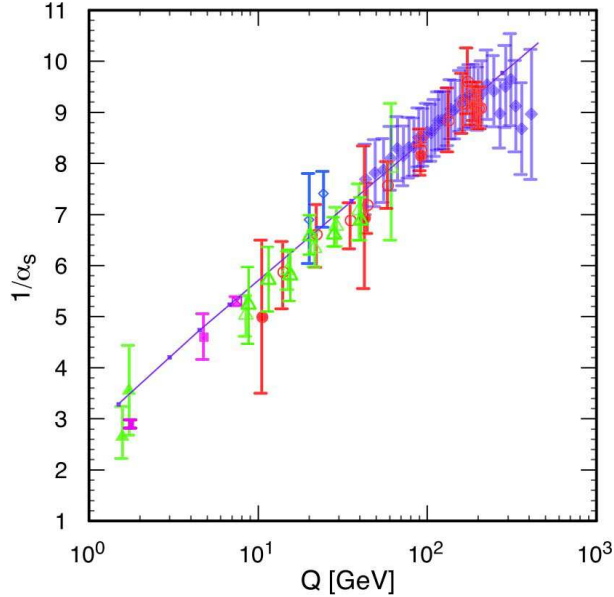}}
\caption{Measurements of  the strong coupling parameter $1/\alpha_s(Q^2)$ as a function of the energy scale $\ln{Q}$. Green triangles: deep inelastic scattering; red circles: $e^+e^-$ annihilations; blue diamonds: hadron collisions; magenta squares: heavy quarkonia.  The curves are the QCD predictions for the combined world average value of  $\alpha_s(M_Z^2)$, in 4-loop approximation and using 3-loop threshold matching at the heavy-quark pole masses $m_c = 1.5\gev$ and $m_b = 4.7\gev$ (from Ref.~{\protect\cite{qcdrl}}, after Ref.~{\protect\cite{Bethke:2006ac}}). Open (filled) symbols indicate next-to-leading-order (next-to-next-to-leading-order) analyses.}
\label{fig:alpharun}
\end{figure}

This essential consequence of QCD has been verified in great detail. I summarize in Figure~\ref{fig:alpharun} a wealth of experimental determinations of $1/\alpha_s$. These exhibit the predicted 
diminution of coupling strength (increase of $1/\alpha_s$) with increasing energy, and agree with the quantitative evolution predicted by the theory. Over a broad range of scales accessible to experiment, perturbative QCD is a powerful and quantitative tool.

The variation of the strong coupling constant with energy scale also means that we can identify a scale with dimensions of mass with a specific value of $1/\alpha_s$. By the phenomenon of \textit{dimensional transmutation,} a theory without any dimensionful parameters reveals a dimensionful scale. A rough way of characterizing the dimensionful parameter $\Lambda_{\mathrm{QCD}}$ is to express the value of the strong coupling at the charm scale as 
\begin{equation}
\frac{1}{\alpha_s(2m_c)} \equiv \frac{27}{6\pi}\ln\left(\frac{2m_c}{\Lambda}\right)\;.
\end{equation}
This insight is important for the understanding of hadron masses. Nonperturbative methods lead us to an expression for the proton mass,
\begin{equation}
M_p = C \cdot \Lambda + \ldots \; ,
\end{equation}
where the dimensionless coefficient $C$ is calculable on the lattice~\cite{Kadoh:2007cv}, the QCD scale parameter $\Lambda$ arises from dimensional transmution, and the small omitted terms are due to quark masses and electromagnetic self energy. Mass arises from a theory that does not rely on any dimensionful parameter. 

An important---one could justifiably say, revolutionary---conclusion is that the nucleon represents a new kind of matter, whose mass is not equal to the sum of the constituent masses. For an isoscalar nucleon, the contribution of quark masses is~\cite{pdg08}
\begin{equation}
3\frac{m_u + m_d}{2} = 10 \pm 2\mev\;.
\end{equation}
Most of the mass of the nucleon is therefore produced, through QCD, by quark confinement, and can be visualized as the kinetic energy of nearly massless quarks or as energy stored up in the gluon field. Thus has lattice QCD made precise the insight that the quark-confinement origin of nucleon mass has explained nearly all the visible mass in the universe.

At this meeting, we have heard reports of new and highly precise determinations of $\alpha_s$ that profit from the combination of beautiful data with heroic perturbative-QCD calculations. I have summarized some of the recent determinations in Table~\ref{tab:alphas}, and refer you to the individual contributions for descriptions of the methods.
\begin{table}[t]
\caption{New and old determinations of the running coupling.\label{tab:alphas}}
\vspace{0.4cm}
\begin{center}
\begin{tabular}{|lccc|}
\hline
Author & Source & Scale & $\alpha_s$  \\
\hline
Malaescu~\cite{Davier:2008sk} & $\tau$ decay & $m_{\tau}$ & $0.344 \pm 0.0086$  \\[6pt]
Gehrmann-De Ridder~\cite{Dissertori:2007xa} & event shapes @ NNLO & $M_Z$ & [$0.1240 \pm 0.0032^a$]  \\
Becher/Schwartz~\cite{Becher:2008cf} & event shapes @ NNLO & $M_Z$ & $0.1172 \pm 0.0021$  \\
Jim\'{e}nez~{\protect\cite{jimen}} &HERA jets & $M_Z$ & $0.1198 \pm 0.0032$  \\[3pt]
Brambilla, et al.~\cite{Brambilla:2007cz} & $\Upsilon(1\mathrm{S})$ radiative decays & $M_Z$ & $0.119^{+0.006}_{-0.005}$\\[3pt]
Bethke~\cite{Bethke:2006ac} & compilation & $M_Z$ & $0.1189 \pm 0.001$  \\
Bethke~\cite{Bethke:2006ac} & jets & $M_Z$ & $0.121 \pm 0.005$  \\
PDG~\cite{pdg08} & compilation & $M_Z$ & $0.1176 \pm 0.002$  \\
LEP EWWG~\cite{ewwg} & global fit & $M_Z$ & $0.1185 \pm 0.0026$  \\
\hline
\end{tabular}
\end{center}
\qquad\qquad$^a$ Awaiting resummation
\end{table}

\section{New Physics Beyond the Standard Model}
Our gathering has also seen several hints of disagreements with the standard model~\cite{Zwicky:2008jp}. Although the putative ``new physics'' does not necessarily implicate QCD, QCD is essential for making the connection between the Lagrangian of the world and the expression of the laws of nature in hadrons. That self-evident point was brought home to me with great force during my summary talk at Heavy Quark 1987~\cite{Quigg:1987by}. I found myself in full rhetorical flight, expounding on the features of $B^0$-$\bar{B}^0$ and $B_s$-$\bar{B}_s$ mixing embodied in the expressions
\begin{eqnarray}
	\left.\frac{\Delta m}{\Gamma}\right)_d & \propto & \; f_{B_d}^2|V_{td}|^2 B_{B_d} {\tau_{B_d}} m_t^2 \nonumber \\
	& & \\
	\left.\frac{\Delta m}{\Gamma}\right)_s & \propto & \; f_{B_s}^2|V_{ts}|^2
B_{B_s} {\tau_{B_s}} m_t^2 \nonumber\; .
\end{eqnarray}
All at once I realized that I could pronounce all the symbols in the equations, but didn't know the value of any one of them! Both experimental and theoretical efforts have brought us very far, and we can now use these oscillation rates to draw important conclusions about the quark-mixing matrix.

The search for new physics in charmed-particle decays has been a topic of interest for many years ~\cite{Hewett:1995aw}. Now we have a specific case to consider, the  puzzle~\cite{Dobrescu:2008er} of the pseudoscalar decay constant $f_{D_s}$, extracted from measurements of the branching fraction
\begin{equation}
B(D_s \to \ell\nu) = \frac{M_{D_s}\tau_{D_s}}{8\pi}f_{D_s}^2\left|G_{\mathrm{F}}V^*_{cs}m_{\ell}\right|^2\left(1 - m_{\ell}^2/M_{D_s}^2\right)^2\;.
\end{equation}
The measured rate for $D_s \to  \ell \nu$ decays, where $\ell$ is a muon or tau, is larger than the standard-model prediction at the $3.8\sigma$ level. Unquenched lattice calculations by the HPQCD Collaboration and UKQCD Collaboration~\cite{Follana:2007uv} give  $f_{D_s} = 241 \pm 3\mev$, with an uncertainty considerably tighter than that estimated by the Fermilab Lattice, MILC, and HPQCD Collaborations~\cite{Aubin:2005ar}, $f_{D_s} = 249 \pm 3 \pm 16\mev$. Experts deem the small claimed uncertainty aggressive, but not manifestly implausible. 
Experiment~\cite{Ecklund:2007zm,Rosner:2008yu,sheldon,Belle:2007ws,Stone:2008gw}, on the other hand, gives  $f_{D_s} = 267.9 \pm 8.2 \pm 3.9\mev$.  At this meeting, Sheldon Stone projected a near-term experimental uncertainty in $f_{D_s}$ of $\approx 7\mev$. The comparison of experiment with lattice QCD is successful for the pseudoscalar decay constant of the $D$-meson.\footnote{For an examination of the state of the lattice art, see~\cite{Gamiz:2008bd}.} On that topic, Benjamin Haas reported~\cite{Becirevic:2007cr} progress on charmed meson decays in lattice QCD. Two unquenched flavors yield $f_D = 200 \pm 22$, to be compared with the HPQCD value of $208 \pm 4\mev$ and the world experimental average, $205.8 \pm 2.5\mev$~\cite{Stone:2008gw}.

What is the outlook for the $D_s$ puzzle? It is important to note first that the experiments are limited by statistics, though the analyses appear robust and have not been impeached in any way. A systematic treatment of the radiative corrections intervene between measurement and a value for the pseudoscalar decay contants is desirable. Other lattice calculations, including some using different dynamical fermions, are highly desirable. In addition to new simulations with $2 + 1$ flavors of dynamical fermions, simulations using $2 + 1 + 1$ may be required to settle misgivings about the influence of the charmed quark. More generally, it is essential to continue testing schemes to remove unphysical copies of fermions. I take the models offered as examples in Ref.~\cite{Dobrescu:2008er} as existence proofs for the kinds of new physics that might resolve the discrepancy between calculation and measurement.

Another hot topic of conversation has been the prospect of new physics in $b \to s$ transitions~\cite{Bona:2008jn}, manifested in the tension between the \textsf{CP} asymmetries in $B^+$ and $B^0$ decays measured by Belle~\cite{:2008zz} and BaBar~\cite{Aubert:2007mj,:2007hh}. The current values of 
$A_{\mathsf{CP}}(B^0 \to \pi^-K^+)$ and $A_{\mathsf{CP}}(B^+ \to \pi^0K^+)$ disagree by $5.3\sigma$~\cite{Peskin:2008zz}.

We heard updates to the ongoing discussion of possible new physics in the anomalous magnetic moment of the muon,  $(g - 2)_{\mu}$~\cite{Miller:2007kk}. Most at issue has been the hadronic contribution to the photon self-energy, for which Benayoun~\cite{Benayoun:2007cu} and H\"{o}cker~\cite{andreash} presented meticulous comparisons of $e^+ e^- \to \hbox{hadrons}$ and $\tau$-decay determinations. Forthcoming information from BaBar promises to be highly informative. Another loose end, the possibility of an unexpectedly large shift in the light-by-light contribution, was analyzed and apparently laid to rest by Vainshtein~\cite{arkady}.

\section{New Physics Within the Standard Model}
Many opportunities to look for new physics (or new manifestations of old physics) also exist within the standard model. Heavy-ion collisions open new realms of QCD~\cite{zakh,cunq,ferre,tyw,kerb}. The energy density of the quark-gluon matter formed at RHIC is $\epsilon \approx 15\gev\fm^{-3}$, and an energy density $3\hbox{ -- }5$ times greater is foreseen for heavy-ion running at the LHC. New phases of matter~\cite{newphases}, \textit{e.g.,} a quark-gluon plasma, a color superconductor, and the famous ``perfect fluid'' are of interest in their own right, and may lead us to ideas useful in other areas, perhaps even electroweak symmetry breaking. Electroweak baryogenesis~\cite{Cohen:1993nk} remains a candidate explanation for the baryon asymmetry of the universe in supersymmetric extensions of the standard model~\cite{carlos}. A bit less familiar is the possible influence of the Wess-Zumino-Witten term~\cite{Wess:1971yu,Witten:1983tw} in the full standard-model Lagrangian, which may show itself in the structure of the decay $\tau^- \to K^- K^+ \pi^-
  \nu_\tau$~\cite{Coan:2004ep}, and in other decays~\cite{Lange:2008dn,Gerard:2005yk,Morales:2008cz}.
  
Two new entries are worth noting: an interaction due to the axial anomaly that induces a $\gamma Z\omega$  vertex~\cite{Harvey:2007ca,Harvey:2007rd}, and an instanton-induced, six-fermion effective Lagrangian to describe decays of the lightest scalar mesons in the diquark--antidiquark picture~\cite{Hooft:2008we}. The anomaly-induced $\epsilon_{\mu\nu\rho\sigma}\omega^\mu Z^\nu F^{\rho\sigma}$ term mediates mediate new interactions between neutrinos and photons at finite baryon density that could have implications for low-energy neutrino scattering in the laboratory and in astrophysical settings. 
In particular, the new interactions might account for the low-energy excess of electromagnetic energy observed by the MiniBooNE experiment~\cite{AguilarArevalo:2007it}.
 An exploratory search for an anomalous energy deposition by neutrinos in a germanium crystal was performed a decade ago in the CERN high-energy neutrino beam~\cite{Castera:1999dr}. That technique might provide a useful probe of the anomaly-induced $\nu\gamma B$ interaction at the lower energies relevant for MiniBooNE. The six-fermion effective Lagrangian mixes tetraquark scalars with corresponding $q\bar{q}$ states and offers an interpretation of the light scalars as $q\bar{q}$/tetraquark mixtures and $Q\bar{Q}$-dominance for the scalar states made of heavy quarks.

With the coming of high-sensitivity experiments that allow us to go over old ground with increased resolving power, I regard the search for new physics within the standard model as very rich terrain for the next few years. A tutorial on the subject at Moriond QCD 2009 would be highly rewarding for theorists and experimentalists alike.

\section{Building Theoretical Apparatus for the LHC}
The construction of the Large Hadron Collider and of the experiments that will soon record 14-TeV proton-proton collisions justly receives great attention---and appreciation---from our community and from the public at large. Less visible, but no less essential to the full exploitation of the opportunities that the LHC will present, has been the work of a devoted cadre of theorists preparing calculational tools and analysis strategies. We were treated to a number of outstanding examples at this meeting.

Important progress on evaluating the differential cross section for Higgs-boson production at next-to-next-to-leading order was reported by Anastasiou~\cite{Anastasiou:2008ik} and Grazzini~\cite{Grazzini:2008fs}. Bernicot~\cite{Bernicot:2008nd} exhibited a calculation of six-photon amplitudes using modern methods. Kosower~ \cite{Berger:2008sj} gave a status report on the automated evaluation of one-loop amplitudes at next-to-leading order, exploiting on-shell methods, and discussed the implementation of gluon amplitudes. Soper~\cite{Soper:2008zp} presented an approach to describing parton showers including quantum interference effects that makes us of the color and spin density matrix and aims to make no approximations other than small-angle and soft-splitting limits. Fuks~\cite{Fuks:2008yp} presented a resummed next-to-leading order calculation of the $Z^\prime$ production cross section, while Sanguinetti~\cite{sanguin} showed an NLO computation of the cross section for $pp \to VV + \hbox{jet}$. Outside the realm of the LHC, Roig~\cite{Roig:2008je} summarized work toward more reliable descriptions of the hadronic structure in $\tau^- \to (KK\pi)^- \nu_\tau$ decays, using modern effective-field-theory techniques.

Much effort is going to refining our knowledge of the parton distribution functions that enter the computation of cross sections in proton-proton collisions. Olness~ \cite{Schienbein:2008ay} described  an ambitions campaign to reduce uncertainties associated with nuclear corrections, while Zoller~\cite{Fiore:2008zq} addressed the role of heavy flavors in the small-$x$ structure functions observed in neutrino-nucleon scattering. Ubiali~\cite{Ubiali:2008yy} advocated the use of neural-net algorithms to escape the (implicit) constraints of parametrizations and to arrive at fully reliable assessments of uncertainties. S\l{}awinska~\cite{Slawinska:2008qy} and Kusina~\cite{Kusina:2008py} gave examples of novel evolution-equation methodologies.

It has long been suspected that the origin of neutrino mass implicates energy scales much higher than can be studied directly~\cite{Seesaw}. Conventional thought accordingly sees the study of neutrino properties as standing apart from collider physics. However, nothing we know excludes the possibility that LHC experiments could have important consequences for the way we think about neutrino physics. Should neutrino masses be set on the 1-TeV scale, related new phenomena should be observed there~\cite{Chen:2006hn}. In particular, within little Higgs models, neutrino masses may be correlated with the existence of doubly charged Higgs bosons, which offer highly characteristic targets for the LHC~\cite{Hektor:2007uu}.

\begin{figure}[tb]
\begin{center}
\centerline{\includegraphics[height=0.4\textheight]{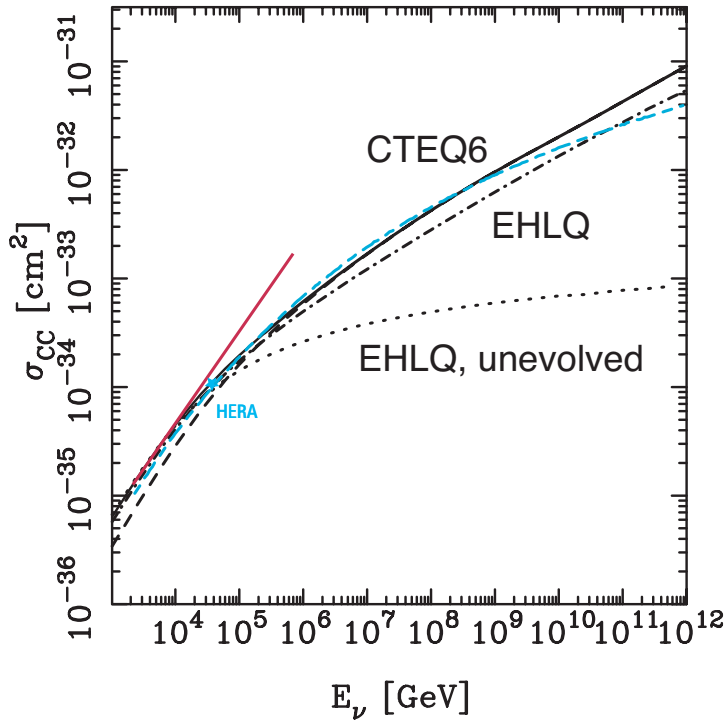}}
\caption{The solid curve shows the charged-current $\nu N$ cross section 
    calculated~\cite{Gandhi:1998ri} using the CTEQ6 parton distributions~\protect\cite{Pumplin:2002vw};
    the dash-dotted line shows the situation in 1986~\cite{Quigg:1986mb}, using Set~2 of 
    the EHLQ parton distributions~\protect\cite{Eichten:1984eu}. The dotted curve shows the energy 
    dependence of the cross section without QCD evolution, i.e., with 
    the EHLQ distributions frozen at $Q^{2} = 5\gev^{2}$. The long-short dashed curve shows the prediction of a structure function that satisfies the Froissart bound for very low $x$ and very large $Q^2$~\cite{Berger:2007ic}. 
\label{fig:uhesigmas}}
\end{center}
\end{figure}

Moreover, from the earliest preparations for supercollider physics, what we have learned about proton structure and parton distribution functions (especially at small values of the momentum fraction $x$) sets our expectations for the ultrahigh-energy neutrino-nucleon interactions to be studied in the great neutrino telescopes. That tradition is alive today, directed particularly toward  the km$^3$ and Pierre Auger observatories~\cite{Armesto:2008cw}.
Figure~\ref{fig:uhesigmas} compares an early calculation, using the 1984
EHLQ structure functions, with the modern CTEQ6 parton
distributions. At the highest energies plotted,
the cross section is about $1.8 \times$ the early estimates, because
today's parton distributions rise more steeply at small $x$ than did
those of two decades ago.  HERA measurements have provided the decisive
new information~\cite{Adloff:2003uh,Chekanov:2002pv,Chekanov:2006ff,CooperSarkar:2007cv}. 
At $10^{12}\gev$, the QCD
enhancement of the small-$x$ parton density has increased the cross
section sixty-fold over the parton-model prediction without evolution.
An ongoing concern, addressed in recent studies~\cite{Anchordoqui:2006ta,Berger:2007ic}, is whether the density of ``wee'' partons becomes so large at relevant values of $Q^2$ that recombination or saturation effects suppress small-$x$ cross sections.
HERA measurements of the charged-current reaction $e p \to \nu +
\hbox{anything}$ at an equivalent lab energy near $40\tev$ observe the
damping due to the $W$-boson propagator and agree with standard-model.
 Reno has given 
a comprehensive review of small-$x$ uncertainties and the possible influence of 
new phenomena on the total cross section~\cite{Reno:2004cx}.

The phenomenon of AdS/CFT correspondence is one of the most startling, and potentially revealing, insights obtained from string theory. It describes uncanny dualities between gauge theories and theories containing gravity. The archetype is an exact equivalence between type IIB string theory compactified on $\mathrm{AdS}^5 \otimes \mathrm{S}_5$ (five-dimensional anti-de Sitter space times the five-sphere) with four-dimensional super-Yang-Mills theory conjectured by Maldacena~\cite{Maldacena:1997re}.
[An anti-de Sitter space refers to a maximally symmetric solution to Einstein's equations with a negative (hence the ``anti'') cosmological constant.]
Holographic QCD~\cite{Sakai:2004cn} is an attempt to find such a gravity dual for quantum chromodynamics, with the initial aim of giving insight into the low-energy properties of hadrons. It reproduces familiar aspects of hadron physics, including the emergence of massless pions as Nambu-Goldstone bosons, a Skyrme~\cite{Skyrme} term, the Wess-Zumino-Witten term~\cite{Wess:1971yu,Witten:1983tw}, and a derivation of the Witten-Veneziano formula for the $\eta^\prime$ mass~\cite{Witten:1979vv,Veneziano:1979ec}.

Intriguing attempts are being made to extract further insights, and even semi-quantitative predictions, from holographic QCD and the AdS/CFT correspondence. In addition to a general review of recent developments~\cite{Zarembo}, we have heard applications to mesons~\cite{Hashimoto} and baryons~\cite{Hong}. Within the framework of Bjorken's hydrodynamics of heavy-ion collisions~\cite{Bjorken:1982qr}, the AdS/CFT correspondence has been employed to try to predict the viscosity of the quark-gluon fluid~\cite{Heller}. There has been interesting technical progress~\cite{Dung} on the suggestion~\cite{Anastasiou:2003kj} that, in 't Hooft's  planar  limit of dimensionally-regulated $\mathcal{N}=4$ super-Yang-Mills theory, higher-loop contributions can be expressed entirely in terms of one-loop amplitudes.

Whether or not the AdS/CFT correspondence ever leads to quantitative predictions or rigorous theorems for QCD, it will be profitable to integrate its results---both conceptual and numerical---into the QCD mainstream for the insights they may provide into strongly coupled theories and for hints about universal behavior. Just as the $1/N_c$ approximation helped explain why the valence-quark approximation could be taken seriously, we may find that even limited results point to new metaphors and new analysis tools for the real quantum chromodynamics.

\section{The Pomeron}
A mysterious entity called the Pomeron has been regularly discussed here in La Thuile, often with the aside that no one under the age of (you choose: 40 \ldots\ 50 \ldots\ 60 \dots) has any idea what it is. The other side of the coin is that colleagues over whatever age you chose do not really know what it is either, but they may at least know what it is supposed to stand for! The term \textit{Pomeron} is intended an an homage to Isaak Yakovlevich Pomeranchuk, a formidable physicist whose wide-ranging contributions to our science are surveyed in~\cite{pomer}. [Moriond QCD stalwart Alexei Kaidalov is the last product of the Pomeranchuk school.] Pomeranchuk stated and proved a famous theorem concerning the asymptotic equality of particle and antiparticle total cross sections (say, $hN$ and $\bar{h}N$)~\cite{pomthm}. When complex-angular-momentum (Regge) analysis of two-body cross sections began to flourish, Gribov designated the highest-lying Regge pole, responsible for what seemed in those early times to be the ``asymptopian'' constancy of the high-energy cross section, as the vacuum singularity. According to Kaidalov, Gell-Mann dubbed it the Pomeranchuk singularity, from which it was abbreviated to Pomeranchukon and then to Pomeron. The Regge intercept of the Pomeron, the location of the pole in the $J$-plane at zero momentum transfer, would be $\alpha_{\mathbb{P}} = 1$ if total cross sections approached constants at high energies.

\begin{figure}[tb]
\centerline{\includegraphics[width=10.0cm]{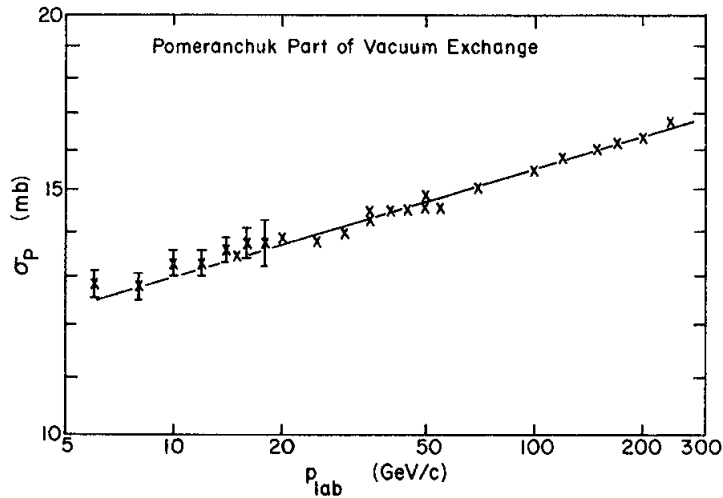}}
\caption{Contribution of the ``$f^0$-free'' combination $\sigma_{\mathbb{P}}$ to meson-nucleon total cross sections [from Ref.~\protect{\cite{Quigg:1975dv}}]. \label{fig:pomeron}}
\end{figure}

The ``vacuum-exchange'' contribution to total cross sections is in fact not governed by a single $J$-plane singularity at intermediate energies. The charge-averaged kaon-nucleon cross sections $\sigma_t(KN) = \cfrac{1}{4}[\sigma_t(K^+p) + \sigma_t(K^+n) + \sigma_t(K^-p) + \sigma_t(K^-n)]$ is a gently decreasing function of beam momentum up to about $p = 30\gev$, after which it rises with increasing energy. Similarly, $\sigma_t(\pi N) = \cfrac{1}{4}[\sigma_t(\pi ^+p) + \sigma_t(\pi ^+n) + \sigma_t(\pi ^-p) + \sigma_t(\pi ^-n)]$ decreases up to about $p = 50\gev$, and then begins to increase. However, the amplitude that corresponds to the combination $\sigma_{\mathbb{P}} \equiv 2\sigma_t(KN) - \sigma_t(\pi N)$, which in quark-model language would be equivalent to $\sigma_t(\phi N)$, eliminates the contribution of an ideally mixed $f^0$ trajectory, and should isolate the contribution of the Pomeranchuk singularity. From the lowest energies, $\sigma_{\mathbb{P}}$ rises according to a power law $p^{\alpha_{\mathbb{P}}-1}$, with $\alpha_{\mathbb{P}} = 1.0755$, as shown in Figure~\ref{fig:pomeron}. This little exercise prefigures the results of comprehensive modern fits to meson-baryon and especially proton-(anti)proton total cross sections initiated by Donnachie and Landshoff~\cite{Donnachie:1992ny}, who found $\alpha_{\mathbb{P}} = 1.0808$.

With the advent of quantum chromodynamics, it was natural to begin searching for a dynamical description of the Pomeron's origin. The idea that the Pomeron somehow emerged from the exchange of color-octet gluons between color-singlet hadrons, first articulated by Francis Low~\cite{Low:1975sv}, has great resonance today. In spite of much productive effort~\cite{qcdpom}, the origin of the Pomeron and the details of its structure are still not clear in every particular.

We have known since the early 1960s that the asymptotic behavior of total cross sections must be bounded by $\sigma_t \ltap (\pi/m_{\pi}^2) \ln^2s$, where $s$ is the square of the c.m.\ energy~\cite{Froissart:1961ux}. The $\alpha_{\mathbb{P}} > 1$ intercept of the Pomeron must therefore be a transitory description of the forward elastic scattering amplitude. There is some ambiguity in the literature about the meaning of ``saturating the Froissart bound.'' I would reserve ``saturating'' for the case in which the coefficient of the $\ln^2s$ term takes on the maximal allowed value (about $62\mb$), but it is often taken to mean the onset of $\ln^2s$ behavior, which indeed provides a consistent representation of the high-energy data~\cite{Block:2005pt}. Here at Moriond, we heard a status  report on how the $\ln^2s$ behavior is realized in QCD~\cite{Avsar:2008dn}; this remains an area of lively discussion.

If the Pomeron can be exchanged between color-singlets, then, in analogy with two-photon physics or the multiperipheral model, two Pomerons can collide and produce collections of particles with net vacuum quantum numbers, isolated by large rapidity gaps from other particle production in the event. A tantalizing possibility is that the Higgs boson might be discovered at the LHC in very quiet events~\cite{Martin:2006fx}.

\section{The Higgs Boson}
Global fits~\cite{ewwg} favor a light Higgs boson, but quantum corrections to precisely measured observable so far only test the Higgs coupling to gauge bosons. Within the standard model, the Higgs boson is the source of fermion masses. We lack direct experimental evidence on this score, so as we search for the Higgs boson at the Tevatron and the Large Hadron Collider, we must be open to the possibility that the  discovery might require nonstandard decay modes. The detection of an ``invisible'' Higgs boson at the LHC could be very challenging indeed~\cite{vanderBij:2008fu}.

The LHC is about more than the search for one particle, of course. Within the context of the 1-TeV scale, we can list a number of big questions about electroweak symmetry breaking \dots\ 
What hides electroweak symmetry?
Is there one Higgs boson? several?
Does the Higgs boson  give mass to fermions, or only to gauge bosons? What sets the fermion masses and mixings? 
How does the Higgs boson interact with itself?
Does the pattern of Higgs-boson decays imply new physics? imply new forms of matter?
What stabilizes $M_H$ on the Fermi scale?
Is Nature supersymmetric?
Is electroweak symmetry breaking controlled by new strong dynamics? by extra dimensions?
And, in view of  the ``LEP paradox''~\cite{Barbieri:2000gf}, can a light Higgs boson coexist with the absence of new phenomena?

\section{Is QCD (Ultraviolet) Complete?}
We have good reasons to think that the electroweak theory is an effective (low-energy) theory: it does not specify the value of the Higgs-boson mass, stability criteria indicate that it must be supplemented below the unification scale unless $134\gev \ltap M_H \ltap 177\gev$ \cite{2loopvacstab}, it does not ensure the separation between the electroweak scale and higher (unification, Planck) scales, and it depends on many apparently arbitrary parameters. QCD exhibits no such structural deficiencies (although the strong \textsf{CP} problem remains to be solved), and so \emph{could,} in principle, be a complete theory up to surpassingly high energies~\cite{Wilczek:1999id} (at least within the renormalization paradigm). That doesn't mean it \emph{must be} the final word: we might encounter new kinds of colored matter beyond the quarks and gluons (and maybe their superpartners), quarks might be composite in an unexpected manner, or the familiar $\mathrm{SU(3)_c}$ gauge symmetry might be the vestige of a larger, spontaneously broken, color symmetry. It is always sound advice to be attentive to surprises.

My personal guess is that the LHC's first surprise in this area will not be a crack in QCD as we know it, but something perhaps buried within QCD (in the nonperturbative or semiperturbative domain) that we have not been clever enough to anticipate. Some unusual structure in a few percent of events that is not suggested by Tevatron experience seems to me a real possibility. Perhaps we will observe high-multiplicity hedgehog events, or sporadic event structures, events containing dozens of small jets, or other manifestations of multiple parton collisions. To recognize the unusual, we need to codify our canonical expectations for the usual---for multiplicities, correlations, and topologies~\cite{Arakelyan:2008cx,Baltz:2007kq}. We also need to devise systematic ways to display and examine the data on particle production. A superb model is the long-ago paper by Ken Wilson that helped launch the systematic study of multiple production at the CERN Intersecting Storage Rings and the Fermilab Tevatron~\cite{kgw}. It is well worth reading (or re-reading) as we prepare for the first LHC collisions~\cite{revol}.

Even in the absence of surprises, the assiduous study of soft collisions and the character of underlying events will pay great dividends for our understanding of multiple production and for the aptness of parton-shower event generators. Thus will it lay a sound foundation to  searches for new physics!

\section{A Modest Proposal}
To help us remain attentive to the possibility of something in the data that we have not anticipated, and to raise our consciousness to phenomena that we might have anticipated in the dim past, I offer a modest proposal for a free-form soir\'{e}e at Moriond QCD 2009: \textit{A Night at the Zoo.} To promote awareness of what nature may have in store for us and to stimulate original thinking, experimental groups will be encouraged to share suggestions of puzzling effects and specimens of ``zoo events.'' Spontaneous theoretical responses are invited. The intention is to hold a though-provoking conversation among a group of friends, with no formal reporting or citations.  As a further inducement to openness, I offer to provide \textit{grappa} to all speakers.

\section*{Remerciements}
Fermilab is operated by Fermi Research Alliance, LLC  under Contract No.~DE-AC02-07CH11359 with the United States Department of Energy. I am grateful to Jim Alexander and the staff of the Cornell Laboratory for Elementary Particle Physics for making Ken Wilson's paper available in digital form. I thank Hans K\"{u}hn and Uli Nierste for warm hospitality in Karlsruhe. Andreas Kronfeld contributed a number of provocative thoughts. It is a pleasure to acknowledge the generous support of the Alexander von Humboldt Foundation. 

\selectlanguage{french}
De la part de nous tous, j'ai grand plaisir de souhaiter un tr\`{e}s grand merci aux personnes qui ont confectionn\'{e} cette splendide semaine de la physique et de l'amiti\'{e} : D'abord \`{a} tous les participants, et surtout aux \'{e}l\`{e}ves, pour leurs excellentes conf\'{e}rences et pour un remarquable engagement; aux gentils organisateurs du comit\'{e} de direction pour le passionnant programme affin\'{e} par leurs soins ; \`{a} nos amies sauvetrices du secretariat pour mille et un miracles ;
au personnel du \textit{Planibel} pour le chaleureux accueil ; et --- \`{a} l'occasion de cette quarante-troisi\`{e}me \'{e}dition des \textit{Rencontres} --- \`{a} Kim et Van pour le bel esprit de Moriond. Bon retour et \`{a} tr\`{e}s bient\^{o}t !

\selectlanguage{english}

\end{document}